\documentclass[a4paper]{jpconf}
\usepackage{graphicx}
\usepackage{amsmath}
\usepackage{cite}
\usepackage{subcaption}
\newcommand{\ttbar}{t\bar{t}}
\newcommand{\mt}{m_t}

\begin{document}

\title{Measurement of the Top Quark Mass From Dileptonic $t\bar{t}$ Decays With 2012 CMS Data}

\author{Richard Nally}

\address{Brown University, on behalf of the CMS Collaboration}

\ead{richard\_nally@brown.edu}

\begin{abstract}
We present a measurement of the top quark mass using 19.7 $\pm$ 0.5 fb$^{-1}$ of $\sqrt{s} = 8$ TeV CMS data. In particular, we study dileptonic $\ttbar$ decays, in which a top-antitop pair decays to a final state containing two electrons or muons. We use the Analytical Matrix Weighting Technique (AMWT), and have performed the first blind top mass measurement at CMS. The mass of the top quark is measured as $m_t = 172.47 \pm 0.17(\text{stat}) \pm 1.40(\text{syst})$ GeV. 
\end{abstract}

\section{Introduction}
As one of the nineteen empirical parameters of the Standard Model, the mass $m_t$ of the top quark is a quantity of extreme experimental and theoretical interest. \ifx First measured with Tevatron data, $m_t$ has been an important topic of research at CMS\cite{Top14015} Recent measurements by CMS\cite{Top14001} and D0\cite{D0ljets}, both analyzing semileptonic $\ttbar$ decays, disagree by nearly 3 GeV, even though both results claim sub-GeV accuracy. Additional measurements of $\mt$ are essential to help resolve this apparent discrepancy. \fi Here we present a measurement of $\mt$ with the full 19.7 $\pm$ 0.5 fb$^{-1}$ of $\sqrt{s} = 8$ TeV CMS data, described in more detail in \cite{PAS}. We analyze dileptonic $\ttbar$ decays, in which both members of a top-antitop pair decay via the CKM-favored $t\to Wb$ decay channel, and then both $W$-bosons decay to lepton-neutrino pairs.\ifx, or symbolically $t\bar{t}\to\ell\nu_\ell b\bar{\ell}\bar{\nu}_\ell \bar{b}$.\fi\ These decays have as their final state observables two jets and two oppositely-charged leptons, as well as two invisible neutrinos. Because of these two neutrinos, it is impossible to directly reconstruct the parent invariant mass; instead, we the Analytical Matrix Weighting Technique (AMWT) to calculate top mass estimators. 

\section{Event Selection and Reconstruction}
We demand two isolated and oppositely charged leptons with $p_T > 20$ GeV, with $\left|\eta\right|<2.4$ for muons and $\left|\eta\right|<2.5$ for electrons, and at least two jets with $p_T > 30$ GeV, with $\left|\eta\right|<2.4$. \ifx Our allowed final state lepton flavors are $ee$, $e\mu,$ and $\mu\mu$.\fi In $ee$ and $\mu\mu$ events, we demand at least 40 GeV of missing transverse energy, and exclude events in the dilepton invariant mass window 76 GeV $<m_{\ell\ell}<$ 106 GeV to remove the dominant $Z$-production background. We reduce Drell-Yan (DY) and QCD contamination by requiring that all events have $m_{\ell\ell}>$ 20 GeV. We split events into categories based on the number of $b$-tagged jets; events are characterized as having either one or two-plus $b$-jets. Our event reconstruction requires the presence of exactly two jets; to implement this, we employ a $b$-jet driven jet-selection, in which the leading-$p_T$ $b$-tagged jets are supplemented by the leading-$p_T$ non-tagged jets.

We analyze 19.7 $\pm$ 0.5 fb$^{-1}$ of data taken by the CMS detector, described in \cite{detector}; this data consists of proton-proton collisions at a center-of-mass energy $\sqrt{s}=$ 8 TeV. We also analyze simulated Monte Carlo (MC) pseudodata. Our signal sample consists of $\ttbar$, and was generated at seven different top mass points. We consider backgrounds corresponding to DY events and single top, W+jets, and diboson (WW, WZ, and WZ) production. The DY contamination in the $ee$ and $\mu\mu$ channels is estimated from data; all other MC samples are normalized to integrated luminosity times NNLO luminosity, calculated for instance in \cite{topXS}.

The details of our event reconstruction are laid out in detail in \cite{PAS,EPJC}, so we will describe our reconstruction only briefly. We construct parent particle mass estimators indirectly. For each event, we loop a top mass hypothesis between 100 and 600 GeV, in increments of 1 GeV, and assign to each hypothesis a weight motivated by the theoretical results of \cite{sonnenschein1,sonnenschein2,sonnenschein3}. Each event is reconstructed 500 times with jet momenta drawn from Gaussians centered at the measured value; the weights of all masses in each of these 500 reconstructions are averaged, and the mass with the highest average sum weight is selected as the top mass estimator, known as the AMWT mass, for that event. We finally impose a cut that all events must have an AMWT mass between 100 and 400 GeV.


The number of events passing all cuts is shown in Table 1. In Figure 1, we plot the distribution of AMWT masses in all data and pseudodata events passing cuts. 

\begin{table}[htp]
\begin{center}
\caption{Total number of data and MC events used in the fit.}
\begin{tabular}{l|cc}
Sample 	& 1 $b$-tag, all leptons	& $\geq2$ $b$-tags, all leptons\\
\hline\hline
$t\bar{t}$	& 23420$\pm$920	& 36400$\pm$1400\\
$W\rightarrow\ell\nu$	&    57$\pm$21	&    20$\pm$14\\
Diboson	&   279$\pm$15	&    61$\pm$4\\
Single Top	&  1348$\pm$95	&  1044$\pm$75\\
Drell-Yan	&  3210$\pm$960	&   620$\pm$180\\
\hline
Total MC	& 28300$\pm$1300	& 38100$\pm$1400\\
\hline\hline
Data	& 27387	& 38705\\
\end{tabular}
\end{center}
\end{table}

\begin{figure}[htb!]
\centering
        \includegraphics[scale=.35]{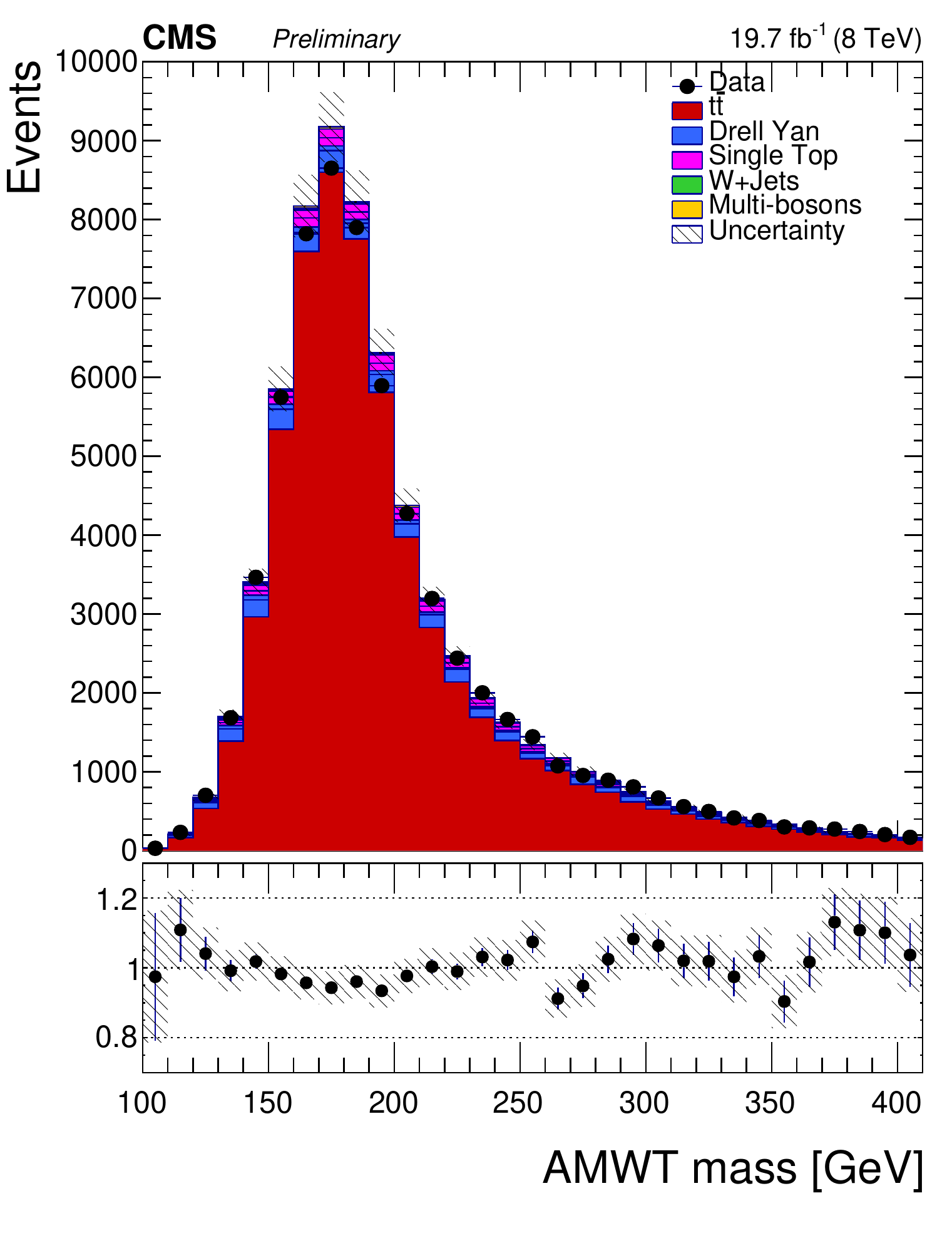}
        \caption{Distribution of the peak mass observed in data and in the simulation generated assuming a top mass of 172.5 GeV. The vertical bars show the statistical uncertainty and the hatched bands show the statistical and systematic uncertainties added in quadrature.}
\label{fig:amwtDistro}
\end{figure}

\section{Fit Calibration and Measurement}
The top mass is measured with a binned likelihood fit of the distribution of AMWT masses in data to distributions from MC templates, generated with values of $m_t$ between 166.5 and 178.5 GeV. We perform a simultaneous fit of two categories: one category is for events with exactly one $b$-tagged jet, and the other is for events with at least two $b$-tagged jets. We calculate a likelihood of fit for each template distribution to the data; a parabola is fit to the negative logs of these likelihoods, and the measured mass is the vertex of the fit parabola.

\begin{figure}[htb!]
\begin{subfigure}[b]{.5\textwidth}
\begin{center}
\includegraphics[scale=.25]{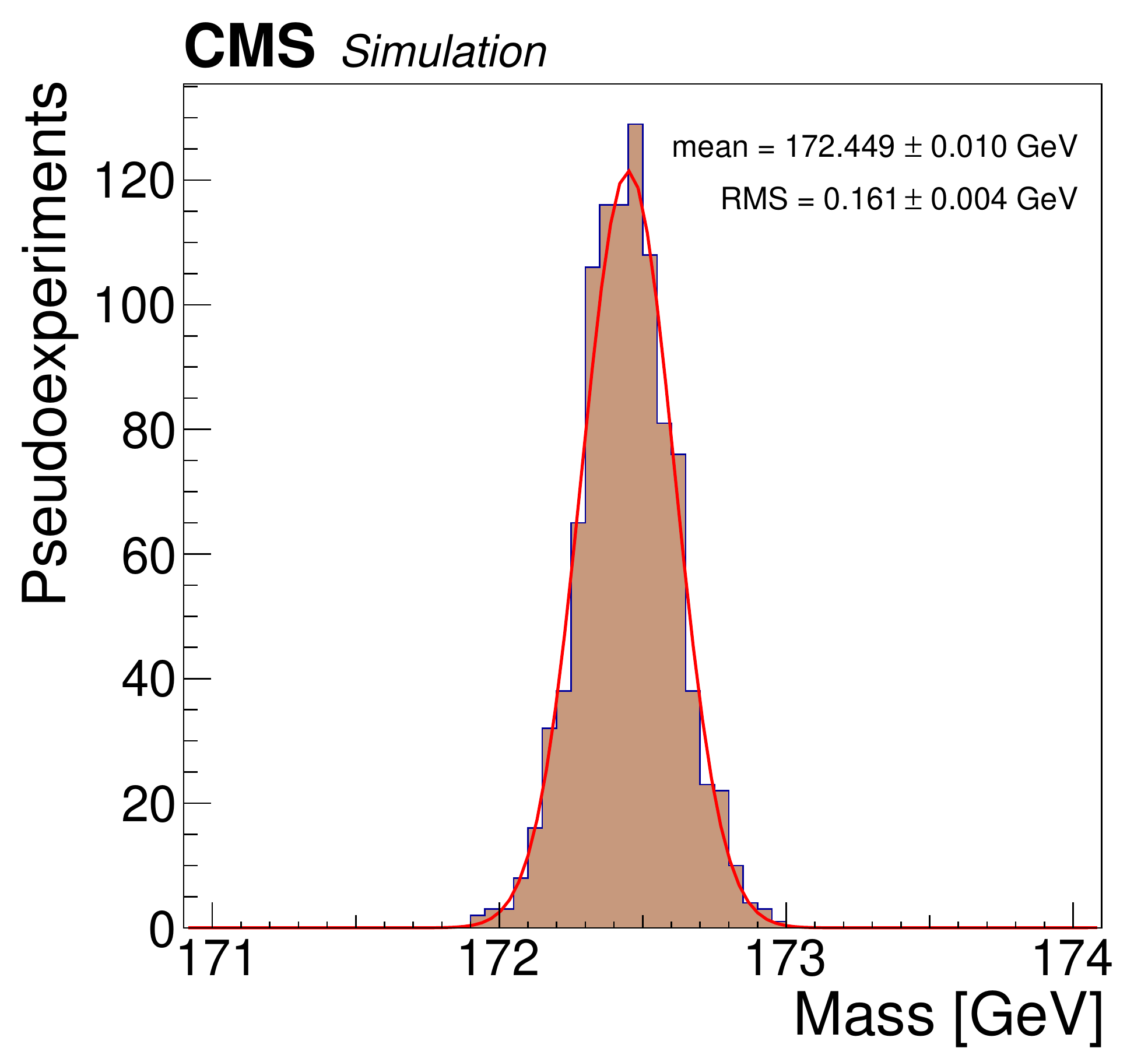}
		\caption{Measured masses of 1000 pseudoexperiments drawn from an MC with $m_t=172.5$ GeV.}
		\label{fig:meanmass}
		\end{center}
	\end{subfigure}
	~
	\begin{subfigure}[b]{.5\textwidth}
	\begin{center}
\includegraphics[scale=.25]{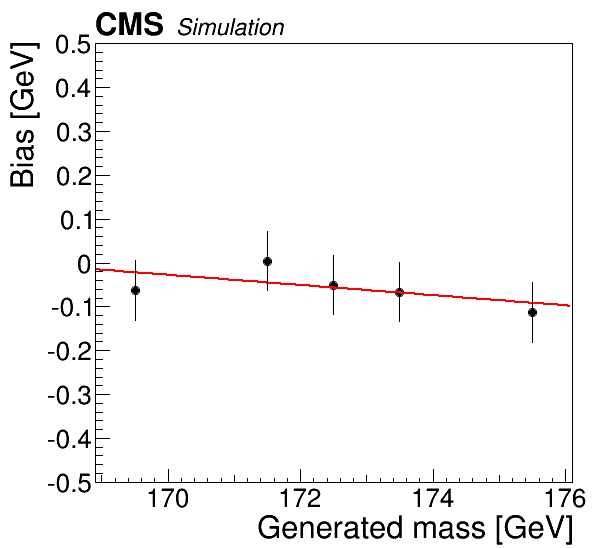}
		\caption{Measured mass versus input mass for pseudoexperiments, with the calibration curve superimposed.}
		\label{fig:calib}
		\end{center}
	\end{subfigure}
	\caption{Results of fit calibration measurements with pseudodata.}
	\label{fig:totalcalib}
	\end{figure}

The fit is calibrated with pseudoexperiments; samples are randomly drawn 1000 times from each of the five central templates (with 168.5 GeV $<\ m_t\ <$ 175.5 GeV), and used as data in a fit. The outcomes of each of these 1000 pseudoexperiments, shown for the nominal MC with $m_t = 172.5$ GeV in Figure \ref{fig:meanmass} are compared to the input MC mass to determine the bias induced by the fit; a line is fit to these biases, shown in Figure \ref{fig:calib}, and is used to correct the bias in the final measurement. 

\begin{figure}[htb!]
\centering
        \includegraphics[scale=.35]{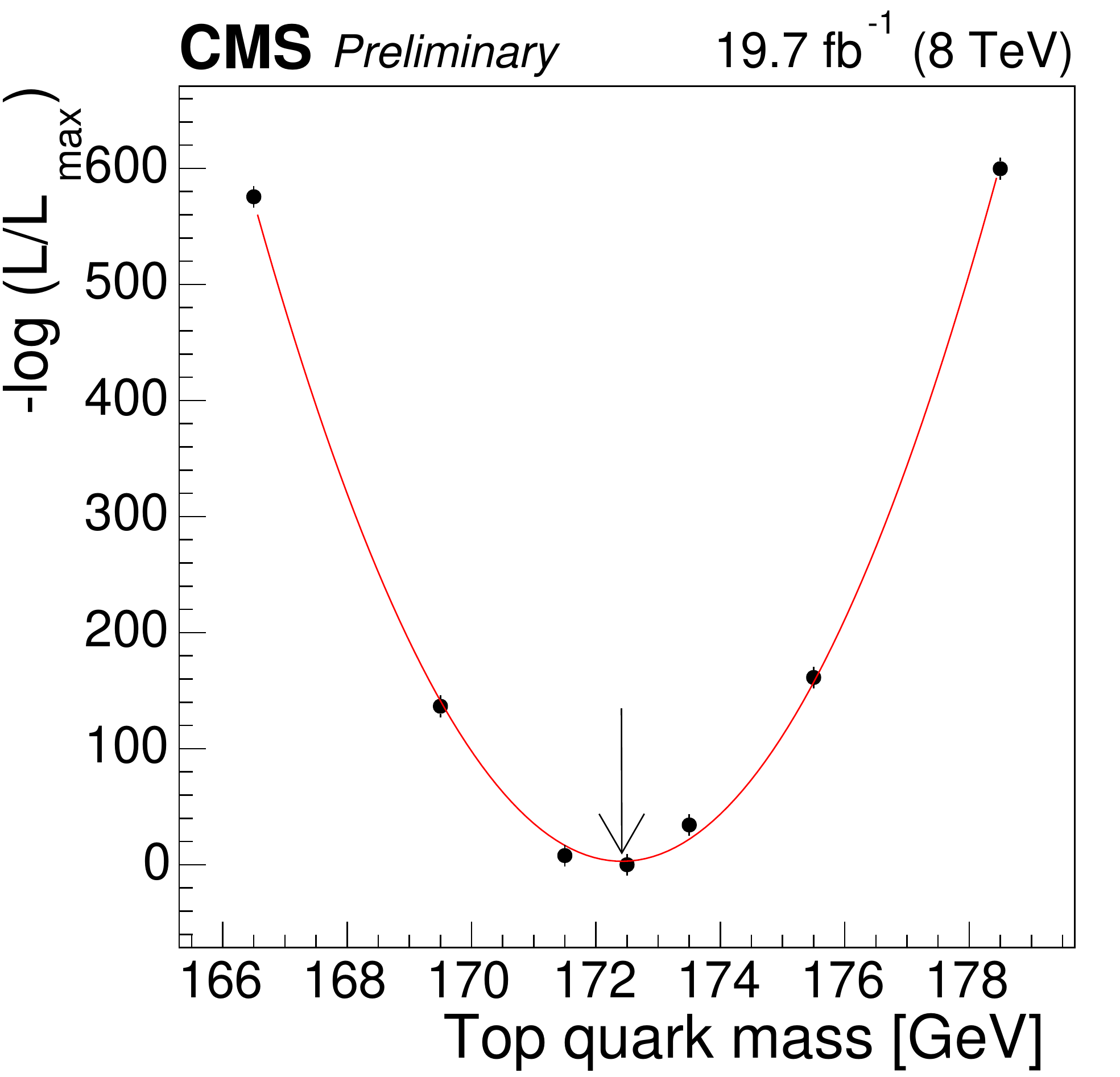}
        \caption{Plot of -log(likelihood) for data. The continuous line is the parabolic fit to the points. The arrow indicates the uncalibrated measured mass of 172.418 GeV.}
\label{fig:fit}
\end{figure}

The fit to data is shown in Figure \ref{fig:fit}. The top mass is measured to be $m_t$ = 172.418 $\pm$ 0.175 (stat.) GeV. After accounting for the fit bias, we measure $m_t$ = 172.47 $\pm$ 0.17 (stat.) GeV. 

\section{Systematics}
In Table 2, we list the different sources of systematical uncertainty considered in this analysis, and their contributions to the overall statistical uncertainty. Details on these sources can be found in \cite{PAS}. The uncertainty of each source is defined as the difference between the mean masses of 1000 pseudoexperiments with events drawn from modified and unmodified templates.

\begin{table}[htp!]
\begin{center}
\caption{Category breakdown of systematic uncertainties.}
\label{table:Yield2B}
\begin{tabular}{l|c}
Source of uncertainty & $\delta m_t$ (GeV) \\
\hline
Fit calibration                   & 0.03 \\
{$p_T$- and $\eta$-dependent jet energy calibration}     & {0.61} \\
Lepton energy scale               & 0.12  \\
Unclustered missing $p_T$                 & 0.07 \\
Jet energy resolution             & 0.09 \\
b tagging                         & 0.04  \\
Pile-up                            & 0.15 \\
Non-$\ttbar$ background            & 0.02  \\
\hline
Flavor-dependent jet energy scale              & 0.28 \\
{b fragmentation}                   & {0.67} \\
Semi-leptonic b hadron decays     & 0.18  \\
\hline
Parton distribution functions                               &  0.18 \\
{Renormalization and factorization scales} & {0.87} \\
Parton-shower matching threshold          & 0.13 \\
Matrix-element generator                      &  0.37 \\
\hline
Underlying event                  & 0.04 \\
Color reconnection modeling       & 0.16 \\
\hline\hline
Total                             & 1.40 \\
\end{tabular}
\end{center}
\end{table}
\section{Conclusion}
We have presented the first blinded measurement of the top mass performed by CMS. Using dileptonic $\ttbar$ decays, we have measured the mass of the top quark to be $m_t = 172.47 \pm 0.17\ (\text{stat}) \pm 1.40\ (\text{syst})$ GeV. This result is in good agreement with the latest world combination \cite{worldcombination}.

\end{document}